\newcommand{\om}{\omega}			
\newcommand{\Om}{\Omega}			
\newcommand{\q}{\vec{q}}                              
\newcommand{\x}{\vec{x}}                               
\newcommand{\ex}{\vec{e}_x}

\newcommand{\ez}{\vec{e}_z}
\newcommand{\DD}{{\mathcal D}}                           
\newcommand{\vG}{\vec{G}}                                         

\newcommand{\Gone} { \vec{G}_1}
\newcommand{\Gtwo} { \vec{G}_2}

\newcommand{\Gx} { \vec{ \mathcal G}_x}  
\newcommand{\gbar} {\bar{g}} 
\newcommand{\erre} { r}  

\newcommand{\As} {\hat{A}_s}

\newcommand{\bs} {\hat{b}_s}
\newcommand{\bi} {\hat{b}_i}
\newcommand{\cs} {\hat{c}_s}
\newcommand{\ci} {\hat{c}_i}

\newcommand {\res}  { \vec{ \mathcal R}} 
\newcommand{\w}{\vec{w}}			
\newcommand{\ws}{\vec{w}_s}

\newcommand{\aaa}{\`a\,\,}

\newcommand{\nn}{\nonumber}
\newcommand{\bsub}{\begin{subequations}}
\newcommand{\esub}{\end{subequations}}
\newcommand{\beq}{\begin{equation}}
\newcommand{\eeq}{\end{equation}}
\newcommand{\beqa}{\begin{eqnarray}}
\newcommand{\eeqa}{\end{eqnarray}}
\newcommand{\beql}{\begin{subequations}\begin{eqnarray}}
\newcommand{\eeql}{\end{eqnarray}\end{subequations}}
\documentclass{osa-article}
\journal{oe}
\articletype{Research Article}
\begin{document}

\title{Efficient parametric generation in a nonlinear photonic crystal pumped by a dual beam}

\author{E. Brambilla,\authormark{1} and A. Gatti\authormark{2,1}}

\address{\authormark{1} Dipartimento di Scienza e Alta Tecnologia dell' Universit\`a dell'Insubria, Via Valleggio 11,  Como, Italy\\
\authormark{2} Istituto di Fotonica e Nanotecnologie del CNR, Piazza Leonardo  Da Vinci 32, Milano, Italy}

\email{\authormark{*}opex@osa.org} 



\begin{abstract}
We investigate parametric down-conversion in a hexagonally poled nonlinear photonic crystal,  pumped by a dual pump  with a   transverse modulation that matches the periodicity of the $\chi^{(2)}$ nonlinear grating. A peculiar feature of this  resonant  configuration  is that the two pumps simultaneously generate  photon pairs over an entire branch of modes, via quasi-phasematching  with  both fundamental  vectors of the reciprocal lattice of the nonlinearity. The parametric gain of these modes depends thus coherently on the sum of the two pump amplitudes and can be controlled by varying their relative intensities and phases.  We find that a significant enhancement of the source conversion efficiency, comparable to that of one-dimensionally poled crystals, can be achieved by a dual symmetric pump.   We also show how the four-mode coupling  arising among   shared modes at resonance 
can be tailored by changing the dual pump parameters.
\end{abstract}

\section*{Introduction}
Since their proposal by Berger \cite{Berger1998} and their first realization \cite{broderick2000}, 
$\chi^{(2)}$ nonlinear photonic crystals (NPC) with a two-dimensional poling pattern 
have attracted great interest due to their potential applications in nonlinear optics
\cite{Saltiel2000,Gallo2008,Stensson2014,Lazoul2015,yellas2017,touami2017,touami2017b},
as well as in quantum optics for the generation and engineering of entangled states
\cite{Jin2013,Gong2012,Megidish2013,Gong2016, gatti2018,jedr2018}.  
Considering parametric down-conversion (PDC) from an intense pump beam into twin photons or twin beams of lower frequencies, the vectors of the 2D reciprocal lattice associated to the nonlinear grating  indeed  offer multiple quasi-phasematching (QPM) possibilities not encountered in more conventional one-dimensional structures. Specifically, the source emission spectrum is characterized by special points, 
the so called "shared modes" lying at the interception
of different QPM branches associated to non collinear lattice vectors \cite{Liu2008,Levenius2012,Chen2014,Conforti2014,jedr2018}. In the stimulated regime of PDC, these shared modes are  cross-seeded by two coupled modes simultaneously, with a significant enhancement of the parametric gain \cite{Liu2008,jedr2018}. These three-mode interaction processes involving the shared modes appear therefore in the source far field as localized bright spots against a more diffused background coming from standard two-mode PDC.

 A recent analysis for a hexagonally poled crystal \cite{jedr2018,gatti2018} demonstrated a peculiar spatial resonance, reached by tilting  the  pump angle till its phase modulation matches the periodicity of the poling pattern: this enforces a transition from  three- to four-mode entanglement among shared modes, the latter being dominated by the Golden Ratio of the segment $\phi = (1 +\sqrt{5})/2$. In the 
 stimulated PDC regime,   a sudden   boost of the intensity of shared modes takes place  as the  pump incidence   angle is  tuned  to resonance  \cite{jedr2018}. The quantum aspects of the underlying entangled state have been investigated in \cite{gatti2018}.

In this work we shall     
explore the possibility of modulating the intensity of the pump beam in the transverse plane, 
rather than only its phase, through the use of a dual pump.
Coherent coupling in $\chi^{(2)}$ materials via dual pumping has been investigated both in a bulk crystal \cite{Danielius93}, and in a one-dimensionally poled QPM structure \cite{Tiihonen2006}, with the realization of self-diffraction in the first case, and resonant cascaded four-wave mixing in the second case. 

In the 2D nonlinear photonics crystal we consider here, the addition of a second pump wave increases the complexity of the source  spectrum,
due to the increased QPM opportunities involving the two pump modes and the two vectors of the 
hexagonal reciprocal lattice. We shall focus on the condition of spatial resonance between the pump modulation and the nonlinear pattern, and  demonstrate some 
unique features not encountered in the conventional single pump configuration. 

A first noticeable feature is the existence of an entire branch of modes (a 3D surface in the Fourier domain) in which photon pairs are  down-converted from both pumps simultaneously,  quasi-phase matched  by  two  fundamental vectors of the reciprocal lattice.
We shall show  that the parametric gain of this QPM branch depends on the coherent sum
of the two pump amplitudes $|\alpha_1+\alpha_2|$  and can be controlled by varying their relative intensities and phases.
In particular, the use of a dual symmetrical pump  ($\alpha_1=\alpha_2$) leads to a substantial increase of the gain compared to a single pump beam of the same energy. 
As a result, the dual pump configuration may in principle bring the source conversion efficiency close to the level of 1D poled crystals of the same material, compensating at least partially for the lower effective nonlinear coefficients associated to two-dimensional poling patterns \cite{arie2007}.

We also show that a dual pumping allows a control over the four-mode processes characterizing the spatial
resonance. Maximum enhancement of the shared mode intensity  is obtained with two  symmetric pumps, while the four-mode coupling degenerates into two independent two-mode processes of reduced gain for anti-symmetric pumps ($\alpha_2=-\alpha_1)$. A dual pump with controlled relative phases and intensities can therefore be used to tailor
the multimode coupling  among shared modes and, in the quantum domain, to  engineer interesting quantum states, which will be the subject of a related investigation \cite{prepgatti}. 

\section{The model}
\label{model}
Our description is based on a model similar to that in \cite{gatti2018}. Even if our analysis may apply to various configurations, we focus here on degenerate down-conversion around $\lambda_s=1064\,$nm from a pump at $\lambda_p=532\,$nm, taking place in a hexagonally poled Lithium Tantalate (LiTaO$_3$) slab similar to the one in \cite{Jin2013}.

\begin{figure}
\begin{center}
\includegraphics[scale=0.4]{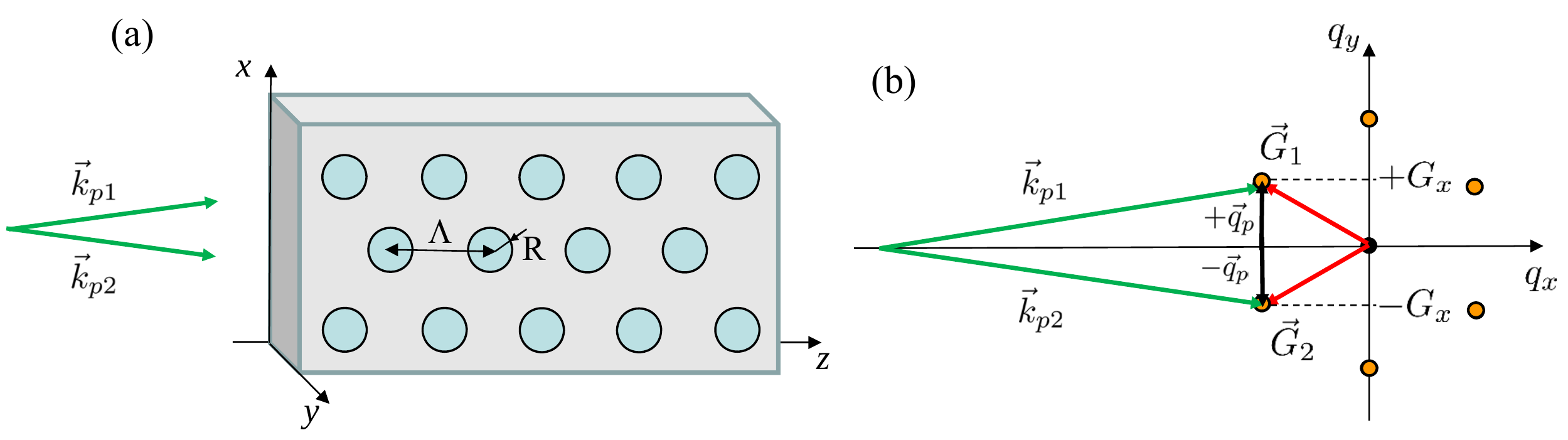}
\caption{  
a) Hexagonally poled $\chi^{(2)} $  crystal pumped by 
 two  waves, symmetrically  tilted
with respect to the  $z$-axis.  
b) A spatial  resonance is achieved by matching
the pump transverse wave-vectors$\pm\q_p$ with the transverse components $\pm G_x\ex$ 
of the lattice vectors $\vG_1$ and $\vG_2$.
}
\label{fig_scheme}
\end{center}
\end{figure}

Figure \ref{fig_scheme} shows the arrangement:  the $\chi^{(2)} $ crystal  is hexagonally  poled in the $(x,z)$ plane orthogonal to the crystal optical axis (the crystal optical axis corresponds to the $y-$axis of the reference frame used in this work, shown in Fig.\;\ref{fig_scheme}a).  We consider  type 0 phase-matching $e\rightarrow e, e$ , where both the  pump and the down-converted field  are  extraordinarily polarized along the optical axis, and propagate mainly  in the $(x,z)$ plane, forming small angles with a mean propagation direction (the $z$-axis in the figure). In particular the pump beam consists of two waves slightly and symmetrically  tilted with respect to the $z-$axis, giving rise to a spatial  transverse  modulation of the pumping of the medium. When the transverse modulation of the pump matches the one of  the nonlinear pattern (Fig.\;\ref{fig_scheme}(b)), a condition of spatial resonance is achieved. 
In such conditions,  quasi phase-matching at the degenerate wavelength $2\lambda_p$ is achieved when the --  poling period  of the two-dimensional pattern   is 
$\Lambda= \frac{2 \pi} {G_z}= 7.782\,\mu$m at a temperature of $85^\circ$C, 
according to the  Sellmeier relations in \cite{Lim2013}. 
These are the conditions chosen for the numerical simulations of the system that will be shown in the following. 

The two-dimensional hexagonal pattern of the nonlinearity   is described by keeping only the leading order terms of  the Fourier expansion of the nonlinear-susceptibility $d(x,z) $\cite{Berger1998}. 
Denoting with $\Gone= -G_x \ex -G_z \ez$ and $\Gtwo = +G_x \ex -G_z \ez$ the two fundamental vectors of the reciprocal lattice 
which  provide  quasi-phase matching, we have accordingly   
\beq
d(x,z) \simeq e^{-iG_z z}\left[ d_{01} e^{-i G_x x} + d_{10} e^{i G_x x} \right] = 2 d_{01}  e^{-iG_z z} \cos{(G_x x)} 
\eeq
 Considering e.g.  inversion domains shaped as discs  of radius $R=0.28\Lambda$  (Fig.\;\ref{fig1}(a)), one has  $d_{01}=d_{10}=0.32 d_{eff}$ \cite{arie2007}, 
 $d_{eff}$ denoting the effective nonlinear susceptibility of the $\chi^{(2)}$ material.

In the degenerate type 0 case,  the signal and idler modes belong to   the same 
wave-packet of central frequency $\omega_s=\omega_p/2$. The signal and pump envelope operators, denoted by $\hat{A}_s$ and $\hat{A}_p$ respectively, satisfy the following coupled propagation equations in the Fourier domain (see \cite{gatti2018} and \cite{Gatti2003, Brambilla2012})
\bsub
\label{NL}
\begin{align}
\frac{\partial}{\partial z}   \hat{A}_s  (\w_s, z )   &=  \chi \int 
 \frac{d^3 \w_p }  {(2\pi)^{\frac{3}{2}} }  \hat{A}_p(\w_p,z) 
\left[   \hat{A}_s^\dagger(\w_p -\w_s -\Gx, z)  e^{-i \DD(\w_s, \w_p -\w_s -\Gx) z }  
\right. \nn \\ & \left.  \hspace{3cm}   +
\hat{A}_s^\dagger(\w_p -\w_s +\Gx, z)  e^{-i \DD(\w_s,\w_p -\w_s +\Gx) z }  \right]
\label{NLs}\\
\frac{\partial}{\partial z}    \hat{A}_p  (\w_p, z )   &= - \frac{\chi}{2} \int 
 \frac{d^3 \w_s }  {(2\pi)^{\frac{3}{2}} }   \hat{A}_s (\w_s,z) 
   \left[ \hat{A}_s  (  \w_p  -\w_s-\Gx, z)  e^{i \DD(\w_s,  \w_p  -\w_s-\Gx ) z } 
 \right. \nn \\  & \left.  \hspace{3cm}   +
 \hat{A}_s  (  \w_p -\w_s +\Gx, z)  e^{i \DD(\w_s,  \w_p -\w_s +\Gx ) z }  \right]
\label{NLp}
\end{align}
\esub
where $\w_j=(\q,\Om)$, $j=s,p$, denotes  the coordinate in the 3D Fourier space of the $j-$th field , with  $\q=(q_x,q_y)$ being the transverse 
 component of the wave-vector,  and $\Om$ the offset frequency from the  
corresponding carrier frequencies $\om_s$ and $\om_p=2\om_s$.
$\Gx= (G_x, 0,0)$  is a short-hand notation for the $x$-component of the reciprocal lattice vector in the 3D Fourier  space, and
$
\chi
\simeq d_{01}
\sqrt{\frac{ \hbar \omega_p  \omega_s^2   }{ 8\epsilon_0 c^3 n_p n_s^2  }}  
$. 
The two terms at r.h.s of Eqs.(\ref{NLp}) describe all the possible three-photon 
processes $\w_p \leftrightarrow \w_s \,, \w_i=\w_p -\w_s \pm\Gx$  mediated by the lattice vectors $\Gone$ and $\Gtwo$ respectively. Notice that for a given pump mode $\w_p=(\q_p,\Om_p)$, the  energy and transverse momentum conservation imply  that $\w_s+\w_i= \w_p \pm \Gx $, i.e. $\Om_s+\Om_i=\Om_p$, and $\q_s+\q_i=\q_p\pm G_x\ex$, 
the latter condition deriving from the simplifying assumption that the crystal is indefinitely extended in the transverse plane.
On the other hand, since we are considering a crystal of finite length $l_c$ along the $z$-direction, the longitudinal momentum conservation  is less stringent, and is expressed by the  phase-matching functions appearing at the r.h.s. of Eqs.
\eqref{NLs} ,\eqref{NLp}
\begin{align}
 \DD(\w_s,\w_p - \w_s  \pm\Gx) =[k_{sz}(\w_s)+k_{sz}(\w_p - \w_s  \pm\Gx)-k_{pz}(\w_p)+G_z], 
\label{DD12}
\end{align}
where
$k_{jz} (\w_j) = \sqrt{k_j ^2(\w_j)  -q^2}$ is  the $z$-components of the wave-vectors associated to mode $\w_j$, the wave number 
$k_j (\w_j) = \frac{\omega_j +\Om}{c} n_j(\w_j)$ being determined by the linear dispersion relation of the $j$-th wave in the medium.  Notice that since all the fields  are extraordinarily polarized,  
 their refractive index  in principle  depends on the propagation direction through $q_y$: 
$n_j(\w_j)= n_e(\omega_j +\Omega_j, q_y)$. 
However, spatial walk-off is negligible since the fields propagate nearly at $\pi/2$ from the optical axis, and more in general 
Lithium Tantalate is characterized by a very small birefringence \cite{moutzouris2011}. Thus we shall neglect the crystal 
 anisotropy both in analytical calculations and in the numerical simulations, letting 
$
 n_j(\w_j) = n_e(\omega_j +\Omega,q_y=0)
$,
where $n_e$ is the crystal extraordinary refractive index in the lattice plane.


\section{Dual plane-wave pump}
In view of obtaining analytical results, we consider the parametric limit where the pump field in Eqs.\;(\ref{NL}) is treated as a classical coherent field undergoing negligible depletion during propagation.
With respect to the pure phase modulation discussed in \cite{jedr2018,gatti2018},  where a spatial resonance 
was realized by tilting a single pump wave,  
we  assume here that the pump consists of  two  monochromatic plane-waves of frequency  $\omega_p$ , 
symmetrically tilted from 
the $z$-axis in the lattice plane, 
as shown  in Fig.\;\ref{fig_scheme}.  
Thus, our analysis includes the possibility of an intensity modulation of the pump. 
Denoting as $\q_{0p}=\pm q_{0p} \ex $ the pump  transverse wave-vectors, in the direct space 
$A_p(\x,t)=\alpha_1 e^{i q_{0p} x}+ \alpha_2 e^{-iq_{0p} x}$, while in  the Fourier domain 
\beqa
A_p(\q,\Om_p)&=&
\int\frac{dt}{\sqrt{2\pi}}
\int\frac{d\x}{2\pi}e^{i\Omega t-i\q\cdot\x}
A_p(\x,t) \nn
\\
&=&(2\pi)^{3/2}\delta(\Om)\delta(q_y)[\alpha_1\delta(q_x-q_{0p})+\alpha_2\delta(q_x+q_{0p})]
\label{Apump}
\eeqa
Quasi-phasematching is achieved for signal modes belonging to one of the four distinct QPM surfaces in the $\w_s$-space: 
\bsub
\beqa
&&\Sigma_{11}:{\cal D}(\w_s,\w_ {0p} -\w_s +\Gx)=0\;\;\;\;\;[\q_s+\q_i=(q_{0p}+G_x)\ex]\label{S11}\\
&&\Sigma_{12}:{\cal D}(\w_s,\w_ {0p} -\w_s -\Gx)=0\;\;\;\;\;[\q_s+\q_i=(q_{0p}-G_x)\ex]\label{S12}\\
&&\Sigma_{21}:{\cal D}(\w_s,-\w_ {0p}  -\w_s +\Gx)=0\;\;\;\;\;[\q_s+\q_i=-(q_{0p}-G_x)\ex]\label{S21}\\
&&\Sigma_{22}:{\cal D}(\w_s,-\w_ {0p}  -\w_s -\Gx)=0\;\;\;\;\;[\q_s+\q_i=-(q_{0p}+G_x)\ex]\label{S22}
\eeqa
\esub
where $\pm\w_{0p}= (\pm q_{0p} , q_y=0, \Om=0)$  denote the 3D Fourier  components of the two pump modes and the relations within the graph parentheses  
give the transverse momentum conservation associated to each QPM surface.
We focus here on the condition of {\em spatial resonance},  obtained by matching the pump transverse modulation to that of the nonlinear lattice, i.e. setting
\beqa
&  q_{0p}=  G_x  \quad \to 
& A_p(z,\x,t)=\alpha_{1} e^{iG_x x} +  \alpha_{2}  e^{-iG_x x} 
\label{resonance}
\eeqa
In this case  $\Sigma_{12}$ and $\Sigma_{21}$ merge into the single surface
\beqa
\Sigma_{12}, \Sigma_{21} \to \Sigma_0: & & {\cal D}(\w_s, -\w_s ) = 0
\label{PM0}
\eeqa
It is important to stress that photon pairs belonging to $\Sigma_0$ originate simultaneously from both pump modes 
and are quasi-phase matched by both vectors $\Gone$ and $\Gtwo$ of the grating of the nonlinearity.
On the other hand, the two other  QPM surfaces 
\beqa
&&\Sigma_{11}:{\cal D}(\w_s,-\w_s +2 \Gx)
=0\;,  \label{PM11}\\
&&\Sigma_{22}:{\cal D}(\w_s,-\w_s -2\Gx)
=0\, 
\label{PM22}
\eeqa
are populated only by  down-conversion either  from pump 1  with the contribution of the lattice vector $\Gone$  (photon  pairs appearing on  $\Sigma_{11}$ ),   or from pump 2 mediated by the lattice vector $\Gtwo$ (photon  pairs appearing on  $\Sigma_{22}$ ). 
\par
Substituting  Eq.\;(\ref{Apump}) into Eq.\;(\ref{NLs}), one obtains the following propagation equation
\begin{align}
\label{propws}
\frac{\partial  \hat{A}_s }{\partial z}(\w_s)
=
(g_1+ g_2) \hat{A}_s^{\dagger}(-\w_s)e^{-i \DD(\ws, -\w_s)z} & 
+ g_1  \hat{A}_s^{\dagger}(-\ws+2\Gx)e^{-i \DD(\ws, -\w_s + 2\Gx)z}\nn
\\
 +  & g_2  \hat{A}_s^{\dagger}(-\ws-2\Gx)e^{-i \DD(\ws, -\w_s - 2\Gx)z}
\end{align}
where  the complex parameters $g_1 = \chi\alpha_1$ and   $g_2 = \chi\alpha_2$  represent the parametric gain per unit length associated to each pump mode. 
The first term at r.h.s. of Eq.\;(\ref{propws}), proportional to the sum of the two pump amplitudes $\alpha_1+\alpha_2$, account for PDC processes taking place on 
the QPM surface $\Sigma_0$, while the last two terms are associated to the QPM surfaces $\Sigma_{11}$ and $\Sigma_{22}$ respectively. 
The propagation  equation (\ref{propws}) for a given mode $\w_s$  has to be considered together with the propagation  equations for the three coupled modes $\hat{A}_s (-\w_s)$ and $\hat{A}_s(-\w_s \pm 2 \Gx)$ appearing 
 at its right  hand side, obtaining 
in principle an infinite chain of coupled  equations for modes of all harmonic orders $\pm\w_s+n\Gx$ (see \cite{gatti2018} for  details). 
However, by inspecting
which modes are effectively phase-matched to the considered signal mode $\w_s$,  the chain can be  truncated   to  finite sets of equations, obtaining thereby 
coupled equations for either 2-mode or 4-mode processes  that will be discussed in the next sections. 
\begin{figure}
\begin{center}
\includegraphics[scale=0.55]{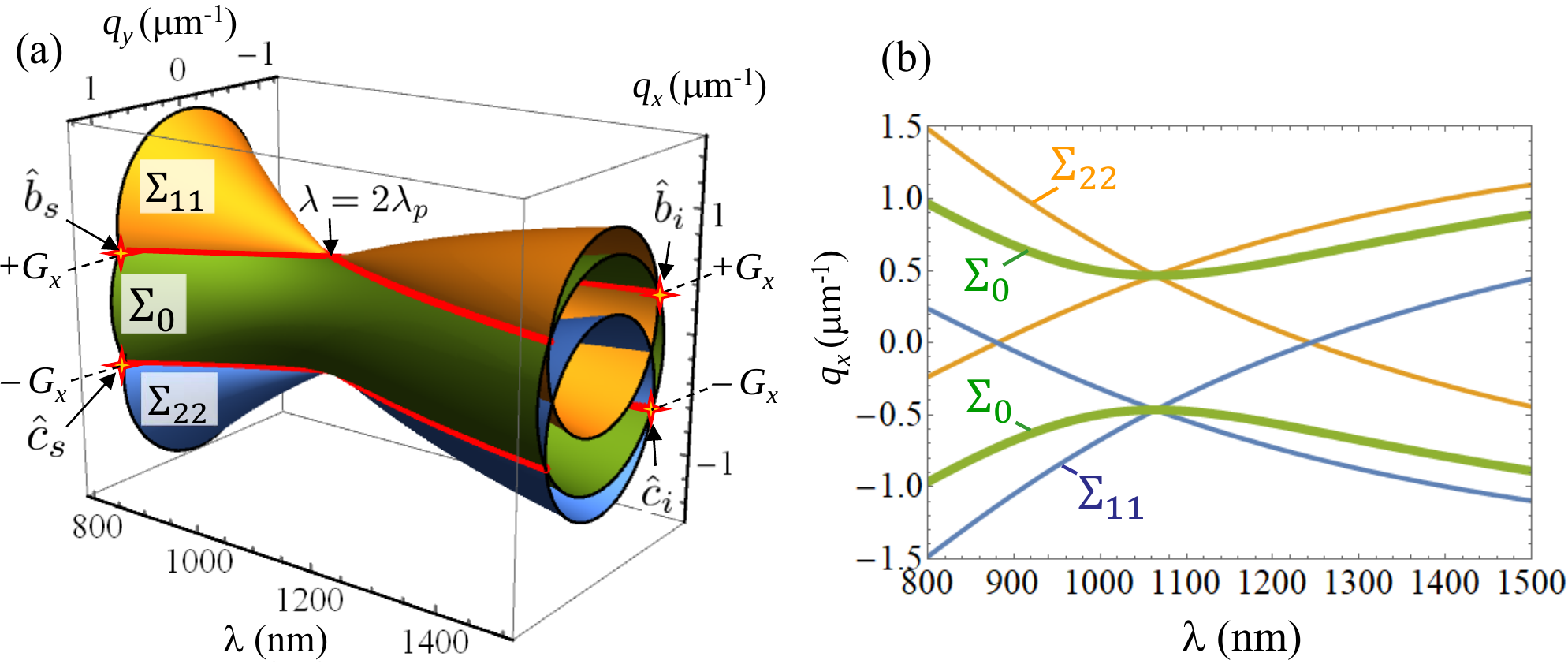}
\end{center}
\caption{Quasi phase-matching in a hexagonally poled LiTaO$_3$ crystal pumped at $532$nm
by a dual pump at spatial resonance with the lattice: $q_{0p} \ex= G_x \ex$, with $G_x= {2\pi}/({\sqrt 3 \Lambda)}=0.466{\rm \mu m}^{-1}$.
(a) QPM surfaces in the 3D Fourier space, and  (b)  its section at $q_y=0$.  
}
\label{fig1}
\end{figure}
\par
Figure \ref{fig1} shows an example of the three QPM surfaces in the 3D Fourier space of the signal modes, obtained by numerically solving  Eqs.\,(\ref{PM0})-(\ref{PM22})  by means of the  Sellmeier formula  reported in \cite{Lim2013}. The period of the poling pattern  is chosen to achieve QPM
at degeneracy  for  the shared signal modes emitted  at $ q_x =\pm G_x, q_y=0 $, as explained  in  App. \ref{appA} [see Eq.\;\eqref{condition}].  In the same appendix we derive 
general  expressions for these QPM surfaces valid within 
the paraxial approximation and not too far from degeneracy.  In particular,  for the chosen crystal parameters,  $\Sigma_{11}$ and $\Sigma_{22}$ are well approximated by the two biconical surfaces  given in  Eqs.\;\eqref{X11} and \eqref{X22}, 
having their  vertexes at $\pm G_x \ex$.
Their projections onto the plane 
$(\lambda, q_x)$ (Fig.\;\ref{fig1}(b)),  are two    X-shaped  curves  symmetrically displaced along the $q_x$ axis , describing  parametric emission  collinear with each  of the two tilted pumps. 
Within the same approximations, $\Sigma_0$ has equation 
$|\q_s\,|\approx(G_x^2+k_s k_s''\Om_s^2)^{1/2}\;,$
corresponding approximately to a wide tube,  centered at $\q=0$,  describing  noncollinear PDC emission at an angle $\theta \approx G_x/k_s $ around the $z-$axis. Notice that  the   $\Sigma_0$ surface includes  the  vertexes of the two cones.  

\subsection{Two-mode processes}
Let us consider signal modes for which only one of the three QPM conditions described by Eqs.\;\eqref{PM0}-\eqref{PM22} is satisfied. These represent the vast majority of modes, with the remarkable exception of modes lying at the intersections of two distinct QPM surfaces, which will be analysed in the next section. Eq.\;\eqref{propws}  reduces then to standard two-mode parametric equations of the form: 
\bsub 
\label{2modes}
\begin{align}
\frac{\partial \hat{A}_s}{\partial z} (\w_s)
= & \gamma \gbar e^{i\phi_1} \hat{A}_s^{\dagger}(\w_i)e^{-i \DD(\ws,\w_i)z}    \\
\frac{\partial \hat{A}_s }{\partial z}  (\w_i)
= & \gamma \gbar   e^{i\phi_1}   \hat{A}_s^{\dagger}(\w_s)e^{-i \DD(\ws,\w_i)z}  
\end{align}
\esub
where $ \w_i = -\w_s $ 
(for modes on the surface $\Sigma_0$ ) 
or $\w_i= -\w_s \pm 2\Gx $
 (on the surfaces $\Sigma_{11}, \Sigma_{22}$) ,  represent   the  twin idler mode coupled to  $\w_s$ . Simple inspection of the different terms at the r.h.s. of Eq.\;(\ref{propws}) allows to determine how the gain of 
these  2-modes processes
compares to the gain from a single pump mode with the same overall  intensity $|\alpha|^2 = |\alpha_1|^2
+|\alpha_2|^2$ . To this end,   we  set
\begin{align}
\gbar&= \sqrt{ |g_1|^2 + |g_2|^2} 
\label{gbar} 
\end{align}
representing  the gain  one would have by concentrating all the energy in a single pump. 
According to the standard solution of the parametric  equations \eqref{2modes}
the number of photons of the phase-matched  modes on each  QPM surface grows as $N \propto \sinh^2 \left(|\gamma| \gbar z \right) $ where, by introducing the complex parameter  $ 
\erre= \frac{g_2}{g_1}$ (without loosing generality we assume that pump 1 is always present, and that  $|g_2 /g_1| \le 1$ ), 
\beq
\gamma =
\begin{cases} 
\frac{g_1+g_2}{\gbar}  = \frac{1+\erre}{ \sqrt{1 + |\erre|^2}} \quad &\text{$\Sigma_0$ }\\
   \frac{g_1}{\gbar}  = \frac{1}{ \sqrt{1 + |\erre|^2}} \quad &\text{$\Sigma_{11}$ }\\
\frac{g_2}{\gbar} =  \frac{\erre}{ \sqrt{1 + |\erre|^2}}  \quad &\text{$\Sigma_{22}$ }
\end{cases} 
\label{gamma}
\eeq
\begin{figure}[b]
\begin{center}
\includegraphics[scale=0.7]{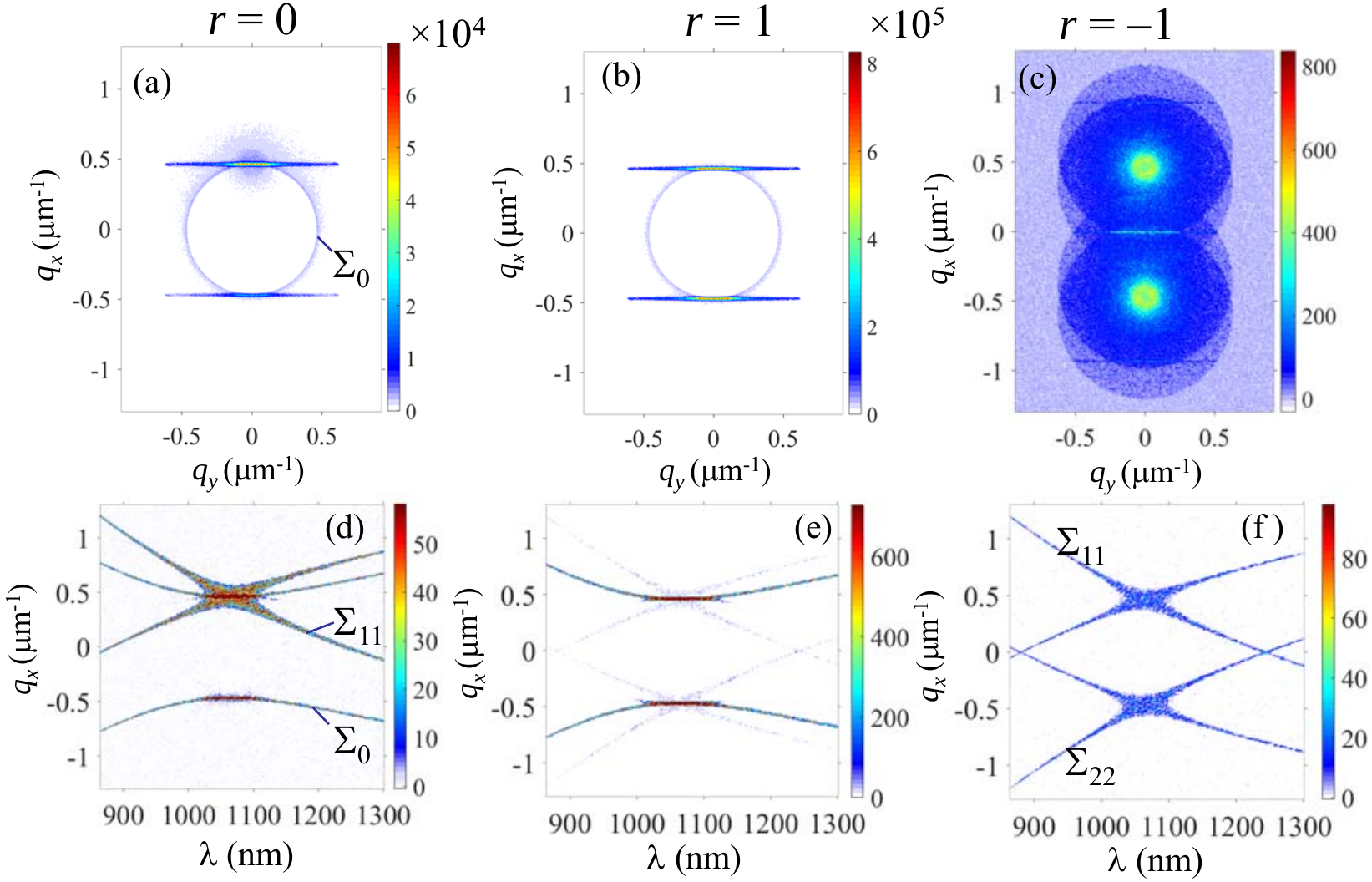}
\caption{ 
Photon-number  distribution in the $(q_x,q_y)$-plane (top) and in the $(\lambda,q_x)$-planes (bottom) for 
a single pump  (a,d),  two symmetric pumps (b,e)  and two
antisymmetric pumps (c,f), from 
numerical simulations  of Eqs.\;\eqref{NL}, in the same NPC of Fig.\;\ref{fig1}.   The pumps are  plane-waves, 
 $\gbar =0.4$ mm$^{-1}$, and   results are shown after 
 $7$mm  of propagation ($\gbar z=2.8$). In (d,e) 
 the scale was truncated to 1\% of the peak value.
For a dual symmetric pump  the  $\Sigma_0$ branch
is significantly more intense than for  a  single pump of equal energy, while it is absent for antisymmetric pumping.
Lines of hot spots at 
$q_x=\pm G_x$ are clearly visible in panels (a) and (b). 
}
\label{fig_numex}
\end{center}
\end{figure}
A striking feature of the spatial resonance (\ref{resonance}) is that the parametric gain of modes belonging to the two-fold degenerate surface $\Sigma_0$  can be controlled by varying the relative intensities and phases of the two tilted pumps. In particular 
\par
\begin{figure}[ht]
\begin{center}
\includegraphics[scale=0.6]{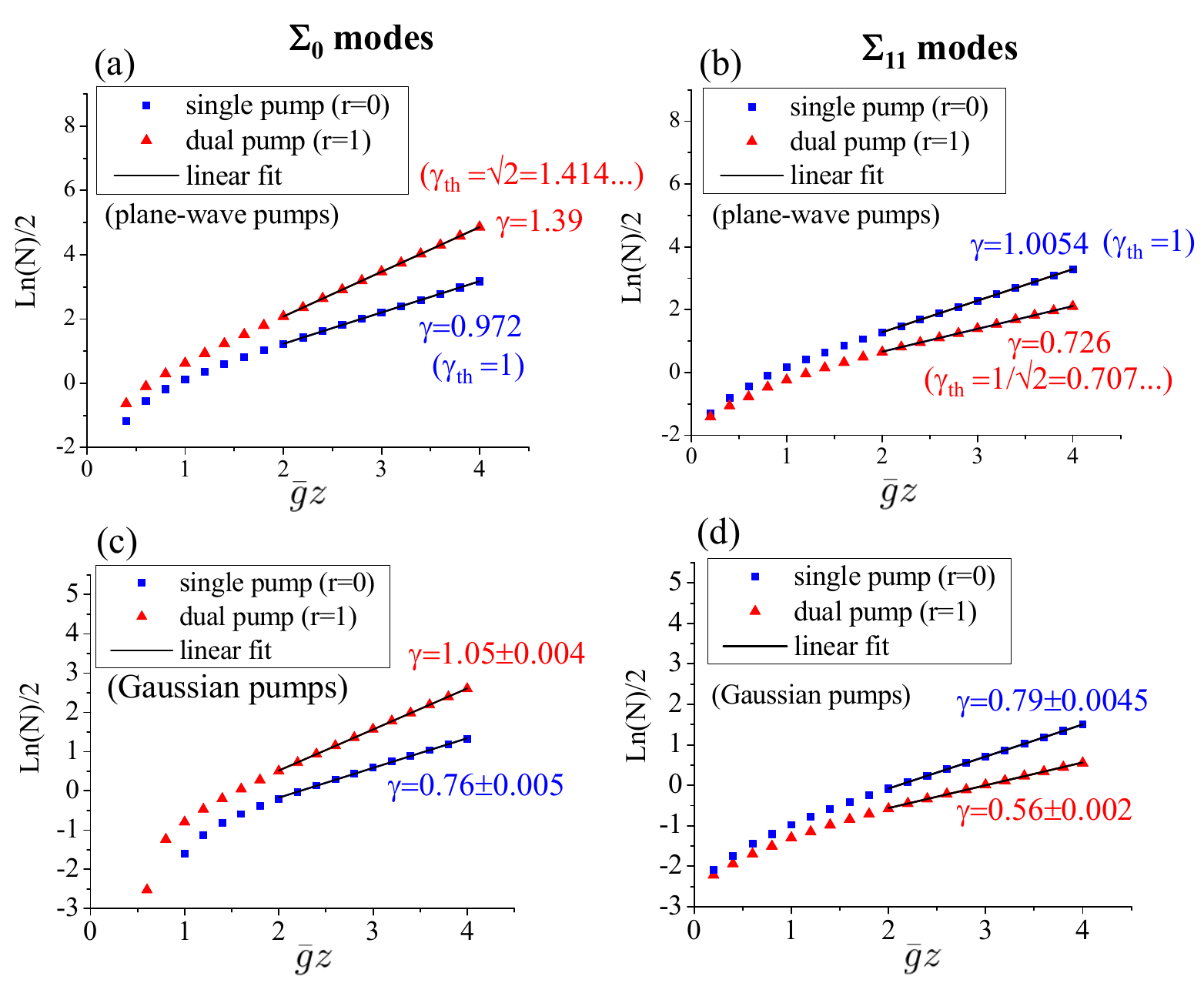}
\caption{ Comparison between the use of a dual symmetric pump (red triangles) and a single pump (blue squares) with the same energy.  The  gain enhancement  factor $\gamma$  is evaluated from numerical simulations of Eqs.\eqref{NL}, 
 The results for  plane-wave pumps in (a,b) are very close to the  $\gamma_{th}$ predicted by the parametric model 
 [Eq.\;\eqref{gamma}].  The lower  panels (c,d)  are obtained for  Gaussian pumps,   of  waists $500\mu$m and $200\mu$m along the $x$ and $y$ axis. Other parameters  as in Fig.\;\ref{fig_numex}
}
\label{fig_gamma2}
\end{center}
\end{figure}

 \noindent 
{\bf i)} 
Maximum gain on $\Sigma_0$ is achieved when the two pumps have the same 
intensity and phase  ($\erre=1$  in  Fig.\;\ref{fig_numex}(b) and \ref{fig_numex}(e)): this symmetric configuration provides a gain  enhancement by a factor $\gamma=\sqrt{2}$ with respect to the use of  a single pump  of equivalent  intensity.    In 	the stimulated regime $\gbar z > 1$,  this leads to a huge increase of the  intensities of the  $\Sigma_0$ modes , as clearly shown by the results in  Fig.\;\ref{fig_gamma2}:    e.g. in Fig\ref{fig_gamma2}(a) for  $\gbar z =4$ the  $\Sigma_0$ modes are roughly 30 times more intense when a dual pump is used instead of a  single pump.
(notice also  the different scales of   Fig.\;\ref{fig_numex}(a) and \ref{fig_numex}(b)). 
We stress that this gain enhancement affects also the PDC emission in the spontaneous regime  $\gbar z \ll 1$, in which  splitting the available energy into a dual pump doubles the intensity of   $\Sigma_0$. 
This occurs because  the resonant structure of the pump  coherently  couples  the processes arising from  the vectors $\Gone $ and $\Gtwo$  of the nonlinear pattern. Thus the use of a dual pump at  resonance  brings a net increase of the efficiency of parametric generation in photonic crystals, over a whole surface of QPM modes.  This is quite different from typical configurations of  PDC in nonlinear photonic crystals involving a single pump wave, where shared modes appear as isolated hot spots  
at the geometrical intersection of different QPM branches 
\cite{Liu2008,jedr2018}. 
Conversely, the 
gain on the side branches $\Sigma_{11}$ and $\Sigma_{22}$ is reduced by a factor $\sqrt{2}$. 
\par \noindent {\bf ii)}
The $\Sigma_0$-modes can be  switched off by taking two antisymmetric pumps, 
with the same intensity and a $\pi$ phase difference  (case $r=-1$ shown in Fig.;\ref{fig_numex}(c) and \ref{fig_numex}(f)). 
Notice that in this case the overall efficiency of quasi-phase-matching with the first order vector of the nonlinear grating is greatly reduced, so that  one should consider also the contribution from higher order harmonics of the grating. The relative weight of those
secondary PDC processes is however strongly dependent on the motif characterizing the poling grating and will not be discussed in this work. 
\par \noindent 
{\bf iii)} 
For a single pump,  corresponding to a pure phase modulation,  the $\Sigma_0$  and  $\Sigma_{11}$ modes  grow with the  same gain $\gbar $ and have the same intensity  ($\erre=0$ case in   Fig.\;\ref{fig_numex}(a) and \ref{fig_numex}(d)). Clearly  the $\Sigma_{22}$ modes
are in the vacuum state because the second pump wave  is absent. This configuration is similar to that  analysed in  \cite{gatti2018} and experimentally demonstrated in  \cite{jedr2018}. 
\par 
The numerical results shown in Figs.\ref{fig_numex} and \ref{fig_gamma2} fully confirm the analysis from the  parametric model.  They are obtained from 3D+1 stochastic simulations of the nonlinear propagation  equations (\ref{NL}).
Numerical integration was  performed using a pseudo-spectral (split-step) method 
in the framework of the Wigner representation, where the field operators are replaced by c-number fields (see e.g. \cite{Drummond97b}).
The input  signal field,  in the vacuum state, is simulated
by Gaussian white noise, while the pump beam is a stationary coherent field
with the transverse spatial modulation  required  by  the resonance condition  Eq.\;(\ref{resonance}). 
The two waves forming the dual pump are either plane-waves or have an elliptical Gaussian profile
with waists $500\mu$m x $200\mu$m along the $x$ and $y$ axis. 
The numerical grid size, 512 $\times$ 256 $\times$ 256 along the $x$, $y$ and temporal axis in the plane-wave pump case, was doubled along the $x$ dimension for the more demanding Gaussian pump case. 
Figure\,\ref{fig_numex} shows the intensity distributions of the down-converted field
after $7$ mm of propagation in the  NPC, comparing the cases $\erre=0$, $\erre=1$ and  $\erre=-1$. The bright lines of 
hot spots at $q_x=\pm G_x$, originate from the 4-mode processes that will be  discussed in the next section. Incidentally, despite the brightness of the hot-spots, for the  chosen  gain $\bar{g} =0. 4 \text{mm}^{-1}$ we verified that   the pump remains almost undepleted .
Fig.\;\ref{fig_gamma2} plots the logarithm of the 
mean number of photons sampled on the QPM branches $\Sigma_{0}$ and $\Sigma_{11}$ 
as a function of $\bar{g}z$ along the crystal, comparing the use  of a single pump 
versus a dual symmetric  pump with the same total energy.  The gain enhancement factors $\gamma$  are estimated by fitting  the numerical data, averaged over portions of the corresponding QPM branches far away from the hot-spots,   with the prediction of  the 2-mode  parametric  model  \eqref{2modes} according to which $N=\sinh^2(\gamma\bar{g}z)$. Precisely,  they are obtained from  linear fits  of the approximated  relation $\log(N)\approx 2\gamma \bar{g}z$  valid for  $\bar{g}z \gg 1$ (black lines). We verified that a nonlinear fit with the function $N=\sinh^2(\gamma\bar{g}z)$ provides similar results.
For plane-wave pumps, 
the   values of $\gamma$  inferred in this way are in good agreement with the analytical predictions (\ref{gamma}). 
For Gaussian pumps,  the effective gain is significantly reduced in all the cases, but 
the ratio of the gains for single and dual pumps are roughly preserved. 
In particular, a dual Gaussian pump provides an enhancement of the gain of the $\Sigma_0$ 
modes by a factor $\sim 1.38$ compared to a single-pump with the same energy, 
very close to the $\sqrt{2}$  factor predicted for plane-wave pumps . The number of generated photons on $\Sigma_0$ 
at the crystal output, for $\bar{g} z=4$, correspondingly  becomes about two orders of magnitude larger.  
\subsection{Shared mode processes}
Modes belonging to the interceptions of two  QPM surfaces  
play a special role, since   two nonlinear processes  concur to the generation of photons,   leading to a  local  enhancement of the gain and to hot spots in the PDC emission \cite{Liu2008,Levenius2012,Chen2014,jedr2018}. 
Here the complexity is further increased by the dual pumping and by the spatial resonance between  the pump and  the nonlinear pattern.
Similarly to  what  happens for  a pure  phase modulation   of the pump \cite{jedr2018,gatti2018}, the spatial  resonance  produces   a  4-mode coupling among shared modes.  As we shall see in the following,   an intensity modulation of the pump can lead to a boost of the hot-spot intensities, and in general permit to tailor the coupling among the 4 modes. 

With a dual pump at spatial resonance with the nonlinear grating [Eq.\;(\ref{resonance})],
the shared modes are determined by the following equations 
\begin{align}
& \DD(\ws,-\ws +2 \Gx)=\DD(\ws,-\ws)  =0 & \text{$\Sigma_0\cap\Sigma_{11}$ }
\label{H01}\\
&  \DD(\ws,-\ws -2 \Gx)=\DD(\ws,-\ws)  =0& \text{$\Sigma_0\cap\Sigma_{22}$ }
\label{H02}\\
&\DD(\ws,-\ws +2 \Gx)=\DD(\ws,\ws -2 \Gx) =0  & \text{$\Sigma_{22}\cap \Sigma_{11}$ }
\label{H12} 
\end{align}
We focus here on  the  intersections  between the central surface $\Sigma_0$  and the two side branches  $\Sigma_{11}$, $\Sigma_{22}$.
As shown in App.\;\ref{appA}, the first equality in  Eqs. \eqref{H01} and \eqref{H02} requires that   
$q_{sx}= +G_x$  and $q_{sx}=- G_x$, respectively.  Requiring in addition that QPM is satisfied on these planes,   one obtains two  continuous lines of shared modes (see the red lines in Fig.\;\ref{fig1}(a)),
that 
for the chosen phase-matching conditions  are characterized by  ${q}_{sy}= \bar {q}_{sy}(\Om_s)\simeq \pm \sqrt{k_s k_s''}\,\Om_s$, as can be  inferred from  Eqs.(\ref{X11},\ref{X22}). 
Thus,  these shared modes exist also at the degenerate frequency $\Omega_s=0$, where they coincide with the vertexes of the  two conical surfaces $\Sigma_{11}$ and $\Sigma_{22}$. 
Conversely,  the shared modes defined by Eq.\;\eqref{H12}, located at $q_x=0$, exist only away from degeneracy and are much weaker (barely visible in Fig.\;3(c)), we shall not discuss them here. 
\par
\begin{figure}
\begin{center}
\includegraphics[scale=0.7]{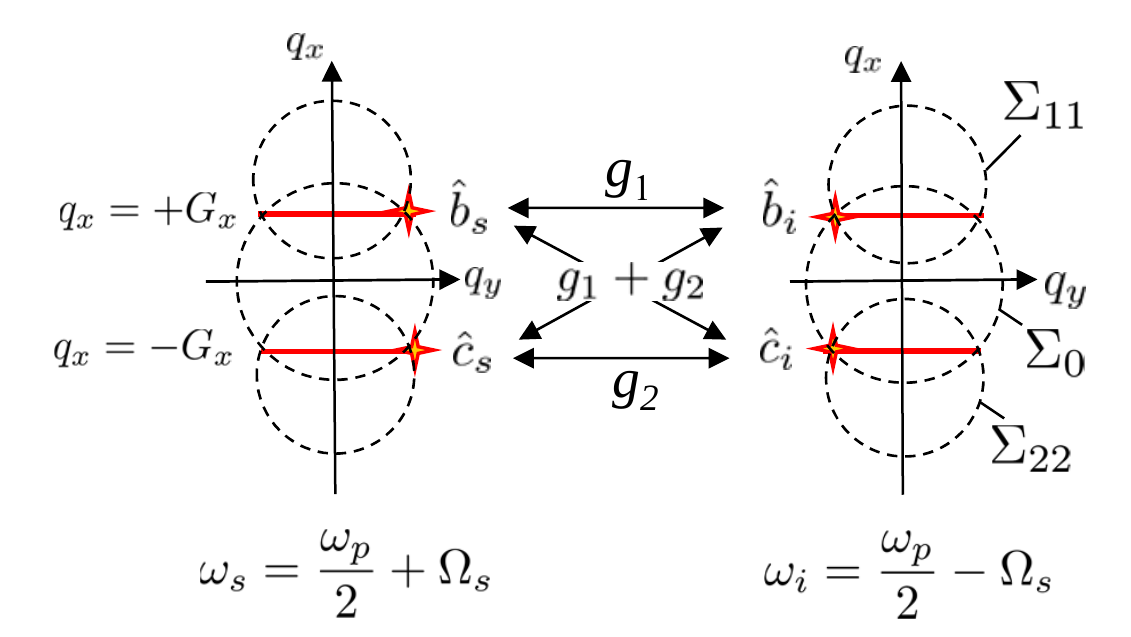}
\caption{Example of four-mode coupling process among shared modes at two conjugate frequencies $\frac{\omega_p}{2}\pm \Om_s$}
\label{fig_4mode}
\end{center}
\end{figure}

 
By inspection of Eq.\;(\ref{propws}), we see that a  given shared signal mode $ \w_s= (+G_x,q_{sy},\Om_s)$  
couples with $(-G_x,-q_{sy},-\Om_s)$  through   both pumps  
and with $(G_x,-q_{sy},-\Om_s)$ through pump 1 only. The 
common phase mismatch function vanishes for $q_{sy}=\bar{q}_{sy}(\Om)$. The third term at $q_x=-3G_x$, proportional to $g_2$, 
is  not phase-matched and can be discarded. A similar reasoning for the other  shared mode $  (-G_x,q_{sy},\Om_s)$  
leads us to conclude that there is a total of four interacting modes, 
for which we shall use the short hand notation 
\begin{align}
&\bs := \As (+G_x, q_{sy},\Om_s) & \bi  :=  \As (+G_x, -q_{sy},- \Om_s)\qquad &\text {modes at $q_x=+G_x$}\\
&\cs  := \As (-G_x, q_{sy}, \Om_s )  & \ci  :=  \As (-G_x, -q_{sy},- \Om_s)\qquad &\text {modes  at $q_x=-G_x$}
\end{align} 
An example of such a quadruplet of coupled modes is shown  in Fig.\;\ref{fig_4mode}.
They satisfy the following coupled equations
\bsub 
\label{par4}
\beqa
&&\frac{d\hat{b}_s}{dz} 
=\left[
 g_1 \hat{b}_i^{\dagger}
+(g_1+g_2)\hat{c}_i^{\dagger}
\right]
e^{-i\bar{D} z}
\\
&&\frac{d\hat{c}_s}{dz}
= \left[
 (g_1+g_2)\hat{b}_i^{\dagger}+g_2 \hat{c}_i^{\dagger}
\right]
e^{-i\bar{D} z}
\\
&&\frac{d\hat{b}^{\dagger}_i}{dz}  
=\left[g_1^*
 \hat{b}_s
+(g_1+g_2)^*\hat{c}_s
\right]
e^{i\bar{D} z}
\\
&&\frac{d\hat{c}^{\dagger}_i}{dz}
=\left[
 (g_1+g_2)^*\hat{b}_s+g_2^* \hat{c}_s
\right]
e^{i\bar{D} z}
\eeqa
\esub
where $\bar D$ is the common phase-mismatch  of the processes. For $\bar D=0$,  
the  eigenvalues  of the 4x4  linear system \eqref{par4} can be readily  found 
as the four real eigenvalues $\pm \Lambda_+, \pm \Lambda_-$, where
\beqa
&&\Lambda_\pm =\left[\frac{2|g_1+g_2|+ |g_1| + |g_2|}{2}   
\pm \frac{1}{2} \sqrt{ \left(|g_1|^2-|g_2|^2\right]^2 +
 4|g_1 +g_2|^4 + 4\left[2   \mathcal{I}m (g_1g_2*)\right]^2 }   \right]^{\frac{1}{2}}  \nn
 \\
&&\to \frac{\gbar }{\sqrt{1+r^2}} \left| \frac{1+r}{2} \pm  \frac{1}{2} \sqrt{5(1+r^2) +6 r} \right| 
=\gbar \times  \begin{cases} 
  \frac{\sqrt{5} \pm 1}{2} \quad & r=0 \\
\frac{3}{\sqrt{2}}, \,  \frac{1}{\sqrt{2}} & r=1 \\
\frac{1}{\sqrt{2}}, \,  \frac{1}{\sqrt{2}} & r=-1 
\end{cases}
 \text{for}\;\erre=\frac{g_2}{g_1}\epsilon \;\mathbb{R}  
\label{LambdaRe}
\eeqa
\begin{figure}[ht]
\begin{center}
\includegraphics[scale=0.5]{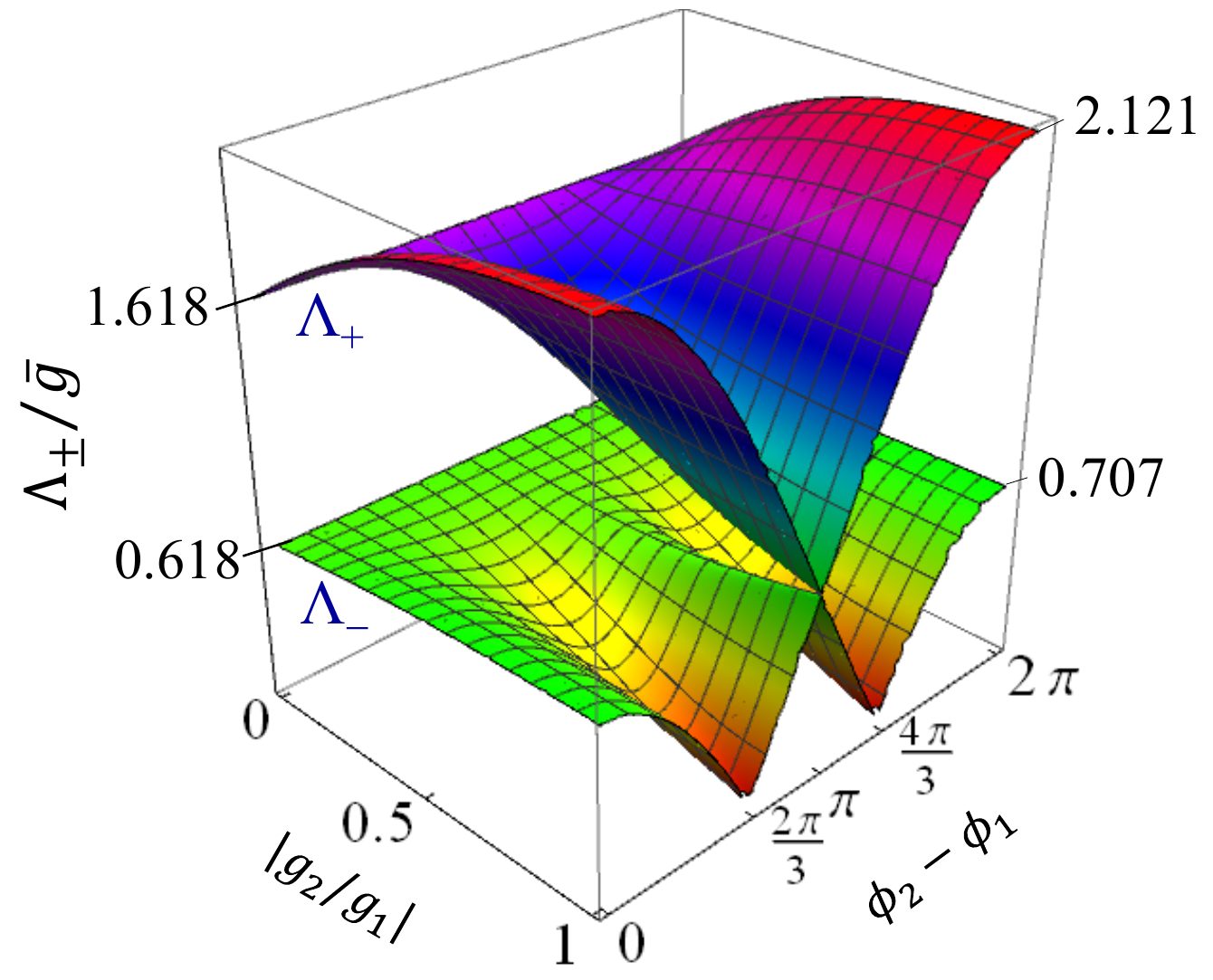}
\caption{ Eigenvalues $\Lambda_+$ (upper surface) and $\Lambda_-$ (lower surface) of the 4-mode propagation equations \eqref{par4}, normalized to $\gbar= \sqrt{|g_1|^2+|g_2|^2}$  as a function of the ratio $|r|=\left| \frac{g_2}{g_1}\right| $  of  the amplitudes of the two pumps and of their  phase difference $\phi_2 -\phi_1$.}
\label{fig_eigenvalues}
\end{center}
\end{figure}
where the second line can be obtained with a little algebra in the simplest case where  $r= \frac{g_2}{g_1} $ is real, i.e. when the two pumps are either in phase or out of phase by $\pi$. 
These eigenvalues are  plotted in Fig.\;\ref{fig_eigenvalues}, as a function of  of the ratio $\left| \frac{g_2}{g_1} \right|$   of  the two complex amplitudes of the pumps and of their  the phase difference $\phi_2-\phi_1$. 
For a single pump at resonance   ($\erre=0$) we thus  retrieve the results  of  \cite{jedr2018,gatti2018}, namely the eigenvalues are $\bar g \Phi$ and $\bar g/ \Phi$, where 
$\Phi= \frac{1+\sqrt{5}}{2}=1.61803..$ is the  Golden ratio. 
When a second pump with nonzero amplitude is injected in the crystal,  the eigenvalues  become modulated with the phase difference between the pumps (Fig.\;\ref{fig_eigenvalues}(b) and \ref{fig_eigenvalues} (c)). The bigger eigenvalue  $\Lambda_+$  reaches its maximum  and minimum values,  $ \, 3\gbar/\sqrt{2} $ and $\gbar/\sqrt{2} $, respectively,  when the two pumps have equal intensities, either in  phase or out of phase by $\pi$.  
\par
\begin{figure}
\begin{center}
\includegraphics[scale=0.6]{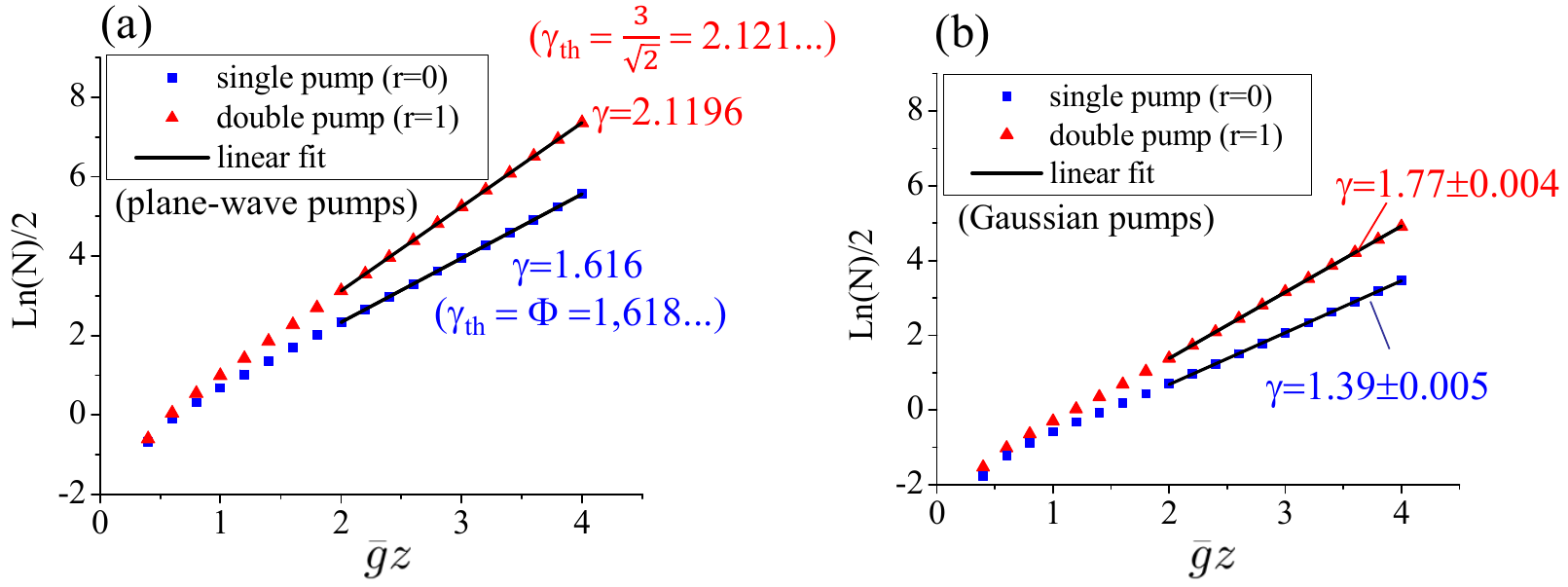}
\caption{Evaluation  of the gain enhancement factor  $\gamma$ in the hot-spots at $q_x=\pm G_x$, from numerical simulations of Eqs.\eqref{NL}. Comparison between
the single pump (blue square) and the dual symmetric pump (red triangles),  for (a) plane-wave pumps, 
(b) Gaussian pumps. The crystal and pump parameters are the same as in Fig.\;\ref{fig_numex} and \ref{fig_gamma2}.}
\label{fig_gamma4}
\end{center}
\end{figure}

 In the high gain regime  $\gbar z \gtrsim 1$,  the number of generated photons is mainly determined  by the big eigenvalue,  which dominates  the exponential rate of growth  of the shared modes along the propagation distance. In this regime,  $\Lambda_+/\bar{g}$ can thus be identified with
the gain enhancement $\gamma $  of the hot-spots involved in  4-mode processes compared to   standard parametric processes 
from a single pump with the same energy. Figure\,.\ref{fig_gamma4} provides instead   an estimation of  the gain enhancement  from numerical simulations of the nonlinear propagation  equations \eqref{NL}, comparing the use of  single and dual pumps, for a given total energy . 
Again, the values obtained for plane-wave pumps in Fig.\;\ref{fig_gamma4}(a) 
turn very close  to the analytical predictions   in Eq.\;(\ref{LambdaRe}) from   the simplified  4-mode model , i.e. $\gamma=\Phi\approx 1.618$ for the single pump ($r=0$) and 
$\gamma=3/\sqrt{2}\approx 2.121$ for the dual pump ($r=1$). 
When considering a Gaussian pump 
profile as in Fig.\;\ref{fig_gamma4}(b), the  effective gains are 
reduced by about $20\%$, but the ratio between the dual pump and the single pump values  is $\simeq 1.27$,  which is 
 close to the plane-wave pump prediction $\frac{3}{\sqrt{2}\Phi }\simeq1.31$.
This last result proves that the enhanced conversion efficiency achieved by using a dual pump with a spatial modulation resonant 
with the NPC grating is a robust feature which affects  not only the whole central branch $\Sigma_0$ but also the  hot spots at $q_{sx}=\pm G_x$.
\begin{figure}
\begin{center}
\includegraphics[scale=0.5]{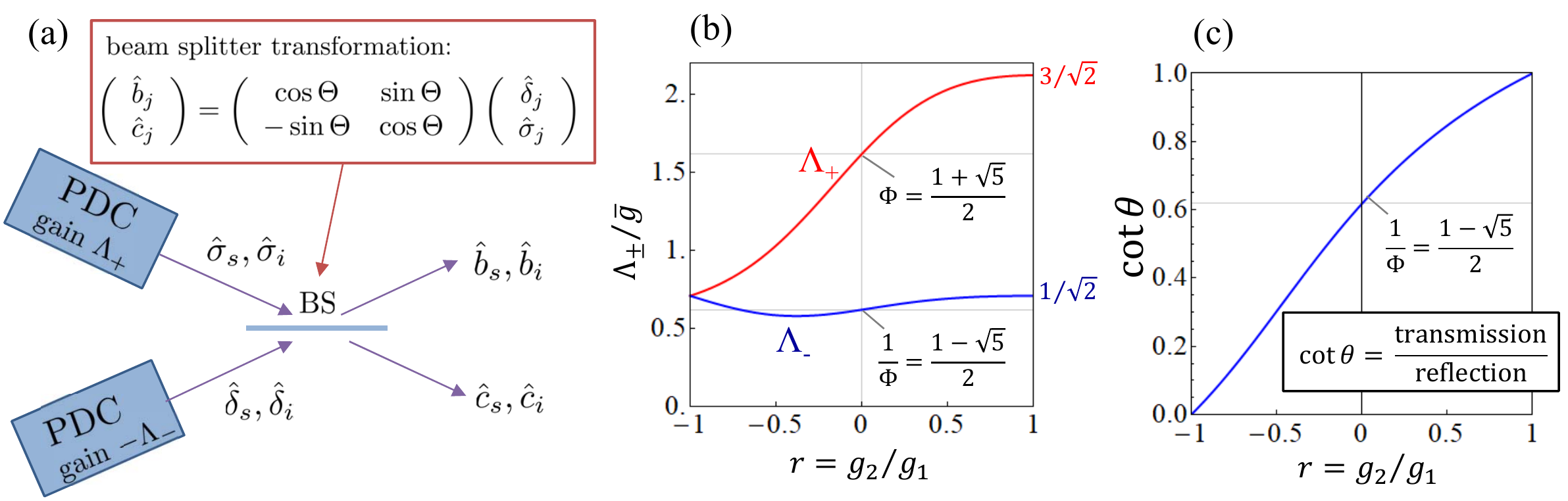}
\caption{(a) For  $\erre=g_2/g_1\, \epsilon\, \mathbb R$ , the 4-mode  process (\ref{par4}) is equivalent to two independent  standard parametric  processes of gains $\Lambda_+$ and $\Lambda_-$ mixed on a beam splitter .
Panel (b) and (c) show the eigenvalues $\Lambda_\pm$ and  the ratio between the transmission and reflection coefficients of the beam-splitter as a function of  $r$ respectively.}
\label{fig_BS}
\end{center}
\end{figure}
\par
In the simplest case  $\erre \, \epsilon \, \mathbb{R}  $, where   the eigenvalues are given by Eq.\;(\ref{LambdaRe}), the four-mode  equations \eqref{par4} can be decoupled by means of a unitary transformation which involve $\hat{b}_s$,$\hat{c}_s$ and 
$\hat{b}_i$,$\hat{c}_i$ separately
\beqa
\label{BS}
\left(
\begin{array}{c}
\hat{\delta}_j \\ \hat{\sigma}_j
\end{array}
\right)
=
\left(
\begin{array}{cc}
\cos\Theta  &- \sin\Theta\\
\sin\Theta & \cos\Theta
\end{array}
\right)
\left(
\begin{array}{c}
\hat{b}_j \\  \hat{c}_j
\end{array}
\right)\;\;\;\;\; j=s,i
\eeqa
with $\cos\Theta=\left(\frac{1}{2}-\frac{1-r}{2\sqrt{5(1+r^2)+6r}}\right)^{1/2}$, 
$\sin\Theta=\left(\frac{1}{2}+\frac{1-r}{2\sqrt{5(1+r^2)+6r}}\right)^{1/2}$.
The new mode operators $\hat{\sigma}_j$ satisfy   standard 2-mode parametric equations 
 with a   gain  per unit length  $\Lambda_+$ 
\beqa
&&\frac{d\hat{\sigma}_{s}}{dz}
=\Lambda_+ 
\, \hat{\sigma}_{i}^{\dagger}
e^{-i\bar{\DD} z}\\
&&\frac{d\hat{\sigma}_{i}}{dz}
=\Lambda_+ 
\, \hat{\sigma}_{s}^{\dagger}
e^{-i\bar{\DD} z}
\eeqa
while the $\hat{\delta}_j$ mode operators satisfy similar equations 
with a comparatively lower gain $\Lambda_-$
\beqa
&&\frac{d\hat{\delta}_{s}}{dz}
=-\Lambda_- 
\, \hat{\delta}_{i}^{\dagger}
e^{-i\bar{\DD} z}\\
&&\frac{d\hat{\delta}_{i}}{dz}
=-\Lambda_-
\, \hat{\delta}_{s}^{\dagger}
e^{-i\bar{\DD} z}
\eeqa
The four-mode coupling process (\ref{par4}) is thus equivalent to 
the outcome of two independent
2-mode parametric processes of gains $\Lambda_+$ and $\Lambda_-$,  followed by  an unbalanced beam splitter that 
 mixes the two pairs of twin output modes 
according to the inverse transformation 
$\hat{b}_j=(\cos\Theta)\hat{\sigma}_j +(\sin\Theta) \hat{\delta}_j$, 
$c_j=(\sin\Theta)\hat{\sigma}_j-(\sin\Theta) \hat{\delta}_j$, $j=i,s$ , as shown in Fig.\ref{fig_BS}(a). Both the gains of the two processes and the reflection and transmission coefficients of the beam splitter can be controlled by varying the relative amplitude and phases of the two pumps, as shown in Fig.\;\ref{fig_BS}(b) and \ref{fig_BS}(c). 
For a single pump  $r=0\, $, we have    $\frac{ \sin\Theta   }{ \cos\Theta}   =\Phi $,   and the beam splitter performs  a Golden ratio partition of the two pairs of twin beams
of gains $\bar{g}\Phi$ and $-\bar{g}/\Phi$, in agreement with the result obtained in \cite{gatti2018}. 
For a  dual symmetical pump $\cos \Theta=\sin\Theta=1/\sqrt{2}$: in this case    two twin modes with gains  $3\bar{g}/\sqrt{2}$ and
$\bar{g}/\sqrt{2}$ are generated and then a mixed on a 50:50 beam splitter.
For an antisymmetric pump,   $\cos\Theta=0$, $\sin\Theta=1$:  indeed in this case  $g_1 + g_2=0$ and  the four-mode process (\ref{par4})
uncouples trivially into two standard parametric  processes 
with equal  gain $\bar{g}/\sqrt{2}$,  which populate only  the  side branches $\Sigma_{11}$ and $\Sigma_{22}$.
\section{Conclusions}
This work  investigated parametric down-conversion in a hexagonally poled nonlinear photonic crystal pumped by  two tilted pumps, forming  a spatial transverse pattern. When the pump pattern matches  the periodicity of the nonlinear grating, we have shown the existence of  a  two-fold degenerate  QPM branch
 of spatio-temporal modes (a surface $\Sigma_0$   in the 3D Fourier space), where 
photon pairs  are generated from  both pump modes , and are  quasi-phase-matched  by both fundamental vectors of the 2D nonlinear grating. \\
In contrast, quasi-phasematching in nonlinear photonic crystals in the  standard  single pump configuration 
generally involves only one lattice vector at a  time, 
except for restricted families of modes of lower dimensionality, the shared modes,
which lie at the interceptions of different QPM branches. 
The two-fold degeneracy of the $\Sigma_0$ branch is thus a distinctive feature of the spatial resonance of the dual pump with the nonlinear grating. As a consequence, its gain per unit propagation length  is proportional to the sum   $|\alpha_1+\alpha_2|$  of  the complex amplitudes of the two pumps, and can be controlled by varying their relative amplitudes and phases.
For two symmetric pumps  ($\alpha_1=\alpha_2$), the down-conversion process is strongly enforced by the pump transverse modulation $\propto \cos(G_x x)$ which oscillates in phase with the nonlinear lattice. Conversely, down-conversion is inhibited for   two  anti-symmetric pumps  $\alpha_2=-\alpha_1$, because the  
  pump modulation $\propto \sin (G_x x)$  is  in quadrature of phase with the nonlinear lattice. 

Specifically,  when comparing  the use of a dual symmetric pump and a single pump with the same total energy,  the conversion efficiency on the $\Sigma_0$ branch simply doubles  in the purely spontaneous PDC regime. However in the stimulated regime,  where the  number of generated photon pairs grows exponentially along the crystal,   the  $\sqrt{2}$  increase of the gain per unit length  leads  to a huge increase of the efficiency. 

We also analysed the  four-mode coupling  characterizing the resonant condition, and we showed that is equivalent to two independent  
parametric processes of different gain generating each a pair of twin beams, followed by an unbalanced beam splitter that mixes the two outcomes
with transmission and reflection coefficients related to the pump amplitudes. 
The present work is mainly devoted to the classical aspects;  the quantum aspects will be discussed in a related work \cite{prepgatti}, showing how
the control of  intensities and phases of the dual pumps   may be used  to tailor the quadri-partite entanglement of the shared modes.

\section{Funding}
This is supported by the Italian 'Ministero dell' Istruzione, dell'Universit\aaa e della Ricerca' (MIUR).

\appendix
\setcounter{equation}{0}
\section{QPM surfaces and shared modes}
\label{appA}
\begin{figure} 
\begin{center}
\includegraphics[scale=0.7]{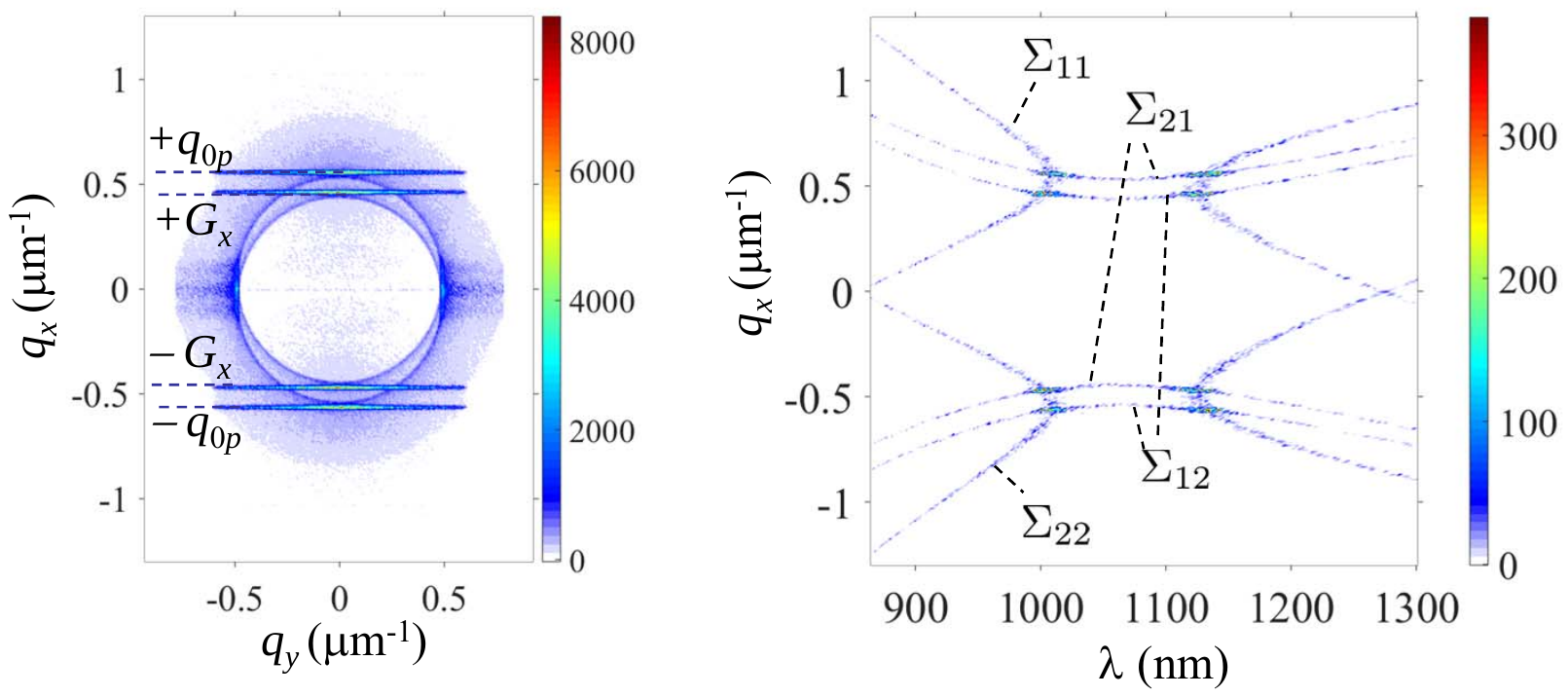}
\caption{ Results of simulations away from  resonance, for  two  symmetric pumps  with  $q_{0p}=1.2\, G_x$. 
(a) Intensity distribution in the $(q_x,q_y)$ plane, showing  four lines of hot spots at 
$q_{x}=\pm q_{0p}$ and $q_x=\pm G_x$. 
(b) Intensity distribution in the  $(\lambda,q_x)$ plane ($q_y=0$). Other  parameters as in Fig.\;\ref{fig_numex}.}
\label{fig_nonres}
\end{center}
\end{figure}

In this appendix  we provide analytical expressions for the QPM surfaces
and for the shared modes at the interceptions of these surfaces. 
We consider the paraxial approximation with the longitudinal components
of the signal wave-vector given by 
\beqa
&&k_{sz}(\ws)\approx k_s(\Om_s)-\frac{q_s^2}{2k_s(\Om_s)}, 
\eeqa
Noticing that the phase-matching functions 
which determine the various QPM surfaces (\ref{S11})-(\ref{S22}) are distinguished only by the resultant  of the pump and lattice transverse wave-vectors: $ \res _{11}= (q_{0p} + G_x) \ex$;   $ \res_{12}= (q_{0p} - G_x) \ex$; 
$ \res_{21}= (-q_{0p} + G_x) \ex$; $ \res_{22}= (-q_{0p} - G_x) \ex$, we can write such functions as
\beqa
\label{Dparax}
& & {\cal D}(\w_s, -\w_s  + \res_{lm} )=
k_{sz}(\w_s)+k_{sz}(- \w_s  + \res_{lm})-k_{pz}(\w_{0p})+G_z \label{A2} \\
& &\quad  \qquad \simeq 
k_{s}(\Om)+k_{s}(-\Om_s)- k_{pz}+G_z
-
\frac{| \res_{lm}|^2}
{2(k_s(\Omega_s)+k_s(-\Omega_s))}
\nn \\
&&\qquad  \qquad
-
\frac{k_s(\Omega_s)+k_s(-\Omega_s)}
{2k_s(-\Omega_s)k_s(-\Omega)}
\left|\q
-\frac{k_s(\Om_s)}{k_s(\Om_s)+k_s(-\Om_s)}
 \res_{lm} 
\right|^2   ,  \qquad  (l,m=1,2)
\label{Dparaxb}
\eeqa
where $k_{pz}$  is  the common z-component of the wavevectors  of the two monochromatic pumps
of wavelength $\lambda_p$ and transverse wavevector  $\pm q_{0p}\ex$. 
According to this result, at a given signal frequency $\Om_s$,  the down-converted field is generated  provided that 
\begin{align}
 D_0(\Omega_s) &= 
k_{s}(\Om)+k_{s}(-\Om_s) -k_{pz} +G_z 
-
\frac{|\res_{lm}|^2}  
{2(k_s(\Omega_s)+k_s(-\Omega_s))} 
\nn \\
&\approx 
2k_s- k_{pz} + G_z   
- \frac{|\res_{lm}|^2}  {4k_s} 
+  k^{''}_s \Omega_s^2  
\ge 0
\label{radii}
\end{align}
where in the second line we made a Taylor series expansion up to second order in $\Omega_s$, valid not too far from degeneracy, and $k^{''}_s =\left.  \frac{d^2 k_s}{d\Omega_s^2} \right|_{\Omega_s=0}$. Provided that the inequality \eqref{radii} is satisfied, then the 
QPM signal modes  lie on 
circumferences  in the $(q_x,q_y)$ plane, 
whose centers 
\beq
\q_{lm} =\frac{k_s(\Om)}{k_s(\Om)+k_s(-\Om)} \res_{lm} \approx \frac{\res_{lm}}{2}
\eeq
are distributed  along the $q_x$-axis, and of radial apertures 
\beq
Q_{lm} =\left[\frac{2k_s(\Om)k_s(-\Om)}{k_s(\Om)+k_s(-\Om)} D_0 (\Omega_s) \right]^{\frac{1}{2}} \approx 
\sqrt{ k_s  D_0 (\Omega_s) }
\label{raggio}
\eeq
In general, in the nonresonant case when $q_{op} \ne G_x$,  one has four distinct QPM branches, corresponding to the four possible resultants of the pump and lattice transverse wavevectors. 
Figure.\,\ref{fig_nonres} shows an example of such four QPM branches,  from a numerical simulation of the evolution equations, for a dual symmetric pump with $q_{op} = 1.2 G_x$. \\
In the resonant case the two surfaces $\Sigma_{12} $ and $\Sigma_{21} $ degenerate into a single surface 
$\Sigma_{0}$ corresponding to a transverse resultant $\res_{12}=\res_{21}=0$ . The other two QPM branches  are characterized by the resultants  $\res_{11}=-\res_{22}=2G_x \ex $. 
We have chosen the poling period in such a way that 
\begin{align}
2k_s- k_{pz} + G_z   
- \frac{|2G_x|^2}  {4k_s} = 0
\label{condition}
\end{align}
Taking into account Eqs.\eqref{radii}-\eqref{raggio},  $\Sigma_{11}$ and $\Sigma_{22}$ are then well approximated 
by two bi- conical surfaces  having their vertexes at  $ q_x = \pm G_x, q_y=0, \Omega=0$: 
\beqa
\label{X11}
&&\Sigma_{11}:|\q_s-G_x \ex |\approx\sqrt{k_s k_s''}|\Om_s|\label{S11appr}\;,\\
&&\Sigma_{22}:|\q_s+ G_x \ex |\approx\sqrt{k_s k_s''}|\Om_s|\label{S22appr}\;,
\label{X22}
\eeqa
which  in the plane $(\Omega_s, q_x)$ appear as two   X's, vertically displaced along $q_x$ of an amount $2G_x$ (See Fig \ref{fig1}b). 
Within the same approximation, $\Sigma_0$ is given by
\beq
\label{S0}
\Sigma_0:|\q_s\,|\approx(G_x^2+k_s k_s''\Om_s^2)^{1/2}\;,
\eeq
which appears like a widely opened tube centered at $\q=0$,  intersecting  the vertexes $\pm G_x  \ex $ of the two conical surfaces as shown in Fig.\;\ref{fig1}(a).  
\par
Let us now determine the interceptions of the  QPM surfaces, corresponding to the shared modes. 
In general,  the QPM modes shared by two surfaces $\Sigma_{lm}$ and $\Sigma_{pq}$ lie on a curve defined by: 
\beq
\Sigma_{lm}\cap\Sigma_{pq} : {\cal D}(\w_s, -\w_s +\res_{lm} )= {\cal D}(\w_s, -\w_s +\res_{pq} )  =0,
\label{shared}
\eeq
Taking into account Eq.\;\eqref{A2}, the first  equality in Eq.\;\eqref{shared} implies that 
\beq
 k_{sz}(- \w_s  + \res_{lm} )=k_{sz}( - \w_s  + \res_{pq}) \quad \to 
 q_{sx} = \frac{\res_{lm} + \res_{pq}}{2}
\eeq
Then for  example the QPM modes shared by $\Sigma_{11}$ and $\Sigma_{12}$ are characterized by the $x$-component of the wave vector $q_{sx}= q_{op}$, while those shared by $\Sigma_{11}$ and $\Sigma_{21 }$  are located at 
$q_{sx}= G_x$. For reasons of symmetry the modes shared by $\Sigma_{22}$ and $\Sigma_{21}$ are located at $q_{sx}= -q_{op}$, while those shared by  $\Sigma_{22}$ and $\Sigma_{12  }$ have $q_{sx}= -G_x$. 
In the plane $(q_x,q_y)$ the shared modes  form four straight lines at $q_x = \pm q_{0p}$ , and $q_x= \pm G_x$, as shown  by the numerical simulations in Fig.\ref{fig_nonres}(a).  The corresponding hot spots have however a much lower intensity than in the resonant case (see Fig.\;\ref{fig_numex}(b)).  
At spatial resonance  $q_{0p}=\pm G_x$,  the four lines  of shared modes
merge into  two ones at   $q_{sx}=\pm G_x$. In the 3D Fourier space, they merge into the two curves of equations
 $ q_{sx}=\pm G_x; q_{sy}= \bar{q}_{sy}(\Om)= \pm\sqrt{k_s k_s''}|\Om_s|$ (red lines in Fig.\;\ref{fig1}(a)). 


 The remaining shared modes belonging to  $\Sigma_{11}\cap\Sigma_{22}$ and $\Sigma_{12}\cap\Sigma_{21}$ 
are  located at $q_{sx}=0$ plane, generally on different curves.  We shall not take these modes into consideration as they lie far from degeneracy.  

\bibliography{biblio.bib}
\end{document}